\documentclass{ws-procs975x65}

\begin{document}
\sloppy
\title{CAN EXTRAGALACTIC FOREGROUNDS EXPLAIN THE LARGE--ANGLE CMB ANOMALIES?}  

\author{Aleksandar Raki\'c$\,^1$, Syksy R\"as\"anen$^2$ and Dominik J. Schwarz$^1$}

\address{$^1$ Fakult\"at f\"ur Physik, Universit\"at Bielefeld, Postfach 100131,
  D-33501 Bielefeld Germany \\ $^2$  CERN Physics Department Theory Unit, CH--1211
  Geneva 23, Switzerland  \\ \email{email: rakic $\rm{\tt at}$ physik $\rm{\tt dot}$
  uni--bielefeld $\rm{\tt dot}$ de, syksy $\rm{\tt dot}$
  rasanen $\rm{\tt at}$ iki $\rm{\tt dot}$ fi,\\
  dschwarz $\rm{\tt at}$ physik $\rm{\tt dot}$ uni--bielefeld $\rm{\tt dot}$ de}}

\begin{abstract}
We address the effect of an extended local foreground on the low--$\ell$
anomalies found in the CMB. Recent X--ray catalogues point us to the
existence of very massive superstructures at the 100~$h^{-1}$Mpc
scale that contribute significantly to the dipole velocity profile. Being highly
non--linear, these structures provide us a natural candidate to leave an imprint on
the CMB sky via a local Rees--Sciama effect. We show that the Rees--Sciama
effect of local foregrounds can induce CMB anisotropy of $\Delta T/T
\sim 10^{-5}$ and we analyse its impact on multipole power as well as the
induced phase pattern on largest angular scales.     
\end{abstract}


\bodymatter

\section{Motivation and Overview}
At largest angular scales which correspond to small multipole moments $\ell$ there
exist puzzling features in the Cosmic Microwave Background (CMB). The near
vanishing of the two--point angular correlation 
function in all wavebands for angular scales between $60^\circ$ and $170^\circ$
is one of the longest known anomalies, already detected in the data of the Cosmic
Background Explorer's Differential Radiometer (COBE--DMR). 
It has been confirmed and persists in the three--year Wilkinson Microwave
Anisotropy Probe data [WMAP(3yr)] 
\cite{wmap3,lambda}. Among the two--point angular correlation functions it has been
shown that none of the almost vanishing cut--sky wavebands matches the full sky
and again neither one of these is in accordance with the best fit
$\Lambda$ cold dark matter ($\Lambda$CDM) model \cite{Copi2}. The disagreement
turned out to be even more
distinctive in the WMAP(3yr) data than in WMAP(1yr) and is unexpected
at $99.97\%$ C.L. for the updated Internal Linear Combination map
[ILC(3yr)]\cite{Copi2}.  

Besides the lack of power, there are a number of remarkable
anomalies regarding the phase relationships of the quadrupole and octopole
within the WMAP data\cite{schwarz,eriksen,tegmark}. In order to be able to make distinct statements
with respect to a phase analysis of multipoles we make use of the \emph{multipole
  vectors formalism}\cite{mvec_form}. 
Looking at quadrupole plus
octopole vectors from WMAP(3yr) the alignment with the equinox (EQX) and with the
ecliptic is found to be unlikely at $99.8\%$ C.L. and $96\%$ C.L. respectively
\cite{Copi2}. The correlation with the dipole direction and with the galactic
plane is found to be odd at $99.7\%$ C.L. and $99\%$ C.L. respectively. 
Moreover from the combined full sky map of $\ell = 2+3$ one infers
that the octopole is quite planar and that the ecliptic 
strongly follows a zero line of the map, leaving the two strongest extrema in
the southern hemisphere and the two weakest in the northern hemisphere.  
Some of these effects are statistically dependent, e.g. given the observed
quadrupole--octopole alignment, the significance of alignment with the galactic
plane is reduced to unremarkable $88\%$ C.L.

These findings support the conclusion that either the Universe as seen by WMAP
is not statistically isotropic on largest scales, or that the observed features
are due to unexpected foregrounds, hidden systematics or new physics
challenging the standard cosmological model. Diverse
attempts for explanation can be found in the literature: 
considering anisotropic or inhomogeneous models [Bianchi family, 
Lema\^itre--Tolman--Bondi (LTB) models]\cite{bianchi,jaffe,alnes,moffat,rakic,tomita}, 
Solar system foreground\cite{frisch}, lensing of the CMB\cite{vale} and moving
foregrounds\cite{cooray}, Sunyaev--Zel'dovich (SZ) effect\cite{abramo,hansen} and  
Rees-Sciama (RS) effect\cite{silk,rakic}, considering a non--trivial topology of
the Universe\cite{luminet,cornish}, considering modifications and refinements of
the standard simplest scenario of
inflation\cite{boyanovsky,campanelli,contaldi1,ferrer,gordon,contaldi2,wolunglee}, 
considering possible phenomenology of loop quantum
gravity\cite{hofmann,tsujikawa}. In this talk we update and expand our previous
work \cite{rakic} in the light of the WMAP(3yr) data release.
 
\section{Local Structures and Rees-Sciama Effect}
Recent X-ray catalogues of
our neighborhood show that a major contribution to the dipole velocity profile
originates from the Shapley Supercluster (SSC) and other density concentrations
at a distance of of around 130--180~$h^{-1}$
Mpc\cite{kocevski04,kocevski05,hudson,lucey}. The SSC is a massive concentration centered
around the object A3558 with a density contrast 
of $\delta \simeq 5$ over the inner 30~$h^{-1}$Mpc region\cite{proust}.
 
We will show that the CMB displays correlations between the dipole 
and higher multipoles after passing through non--linear structures, due 
to the RS effect\cite{reessciama}. The physics of the RS effect is that in the 
non--linear regime of structure formation, the gravitational potential 
changes with time, so photons climb out of a slightly different 
potential well than the one they fell into.
Following ref.\cite{panek} the CMB anisotropy produced by a spherical superstructure
is estimated by the integral of the gravitational potential perturbation $\phi
\simeq \delta M/d$ along the path of the photon: $\Delta T(\theta, \varphi)/T
\simeq \phi \, v_c \,$, where $d$ is the physical size of the structure and
$\delta M$ is the mass excess. Here we assumed a structure collapsing at velocity
$v_c$ and let the evolution time of the structure $t_c$ be the matter
crossing time $d/v_c$ (using $c\equiv 1 \equiv G$). We estimate the typical
collapse velocity from the
energy balance condition $v_c^2 \simeq \phi$ and get: $\Delta T(\theta,
\varphi)/T \sim \phi^{3/2} \sim (\delta M/d)^{3/2}$. We model the
non--linear structure by a spherically symmetric LTB model embedded in a flat
($\Omega = 1$) Friedmann--Robertson--Walker Universe. Substituting the expression
for the mass excess within this model we arrive at\cite{panek}:
\begin{equation}
 \frac{\Delta T(\theta, \varphi)}{T} \sim \left( \frac{\delta \rho}{\rho}
 \right)^{3/2} \left( \frac{d}{t} \right)^3 \; \; ,
\end{equation}  
where $t$ is the cosmic time at which the CMB photons crossed the structure.

Inserting the characteristics of the SSC it follows that {\it a CMB
  anisotropy of $10^{-5}$ due to a local RS effect is reasonable.}
For simplicity we picture the local Universe as a spherically 
symmetric density distribution, with the Local Group (LG) falling towards the core of 
the overdensity at the centre. The line between our location
and the centre defines a preferred direction $\hat{\boldsymbol z}$, which in the 
present case corresponds to the direction of the dipole. This setup exhibits
rotational symmetry w.r.t.~the axis $\hat{\boldsymbol z}$ (neglecting transverse
components of our motion). Consequently, only zonal
harmonics ($m=0$ in the $\hat{\boldsymbol z}$-frame) are generated. Note that
any other effect with axial symmetry would also induce anisotropy only in the
zonal harmonics. 
 
\begin{figure*}[t!]
\epsfig{figure=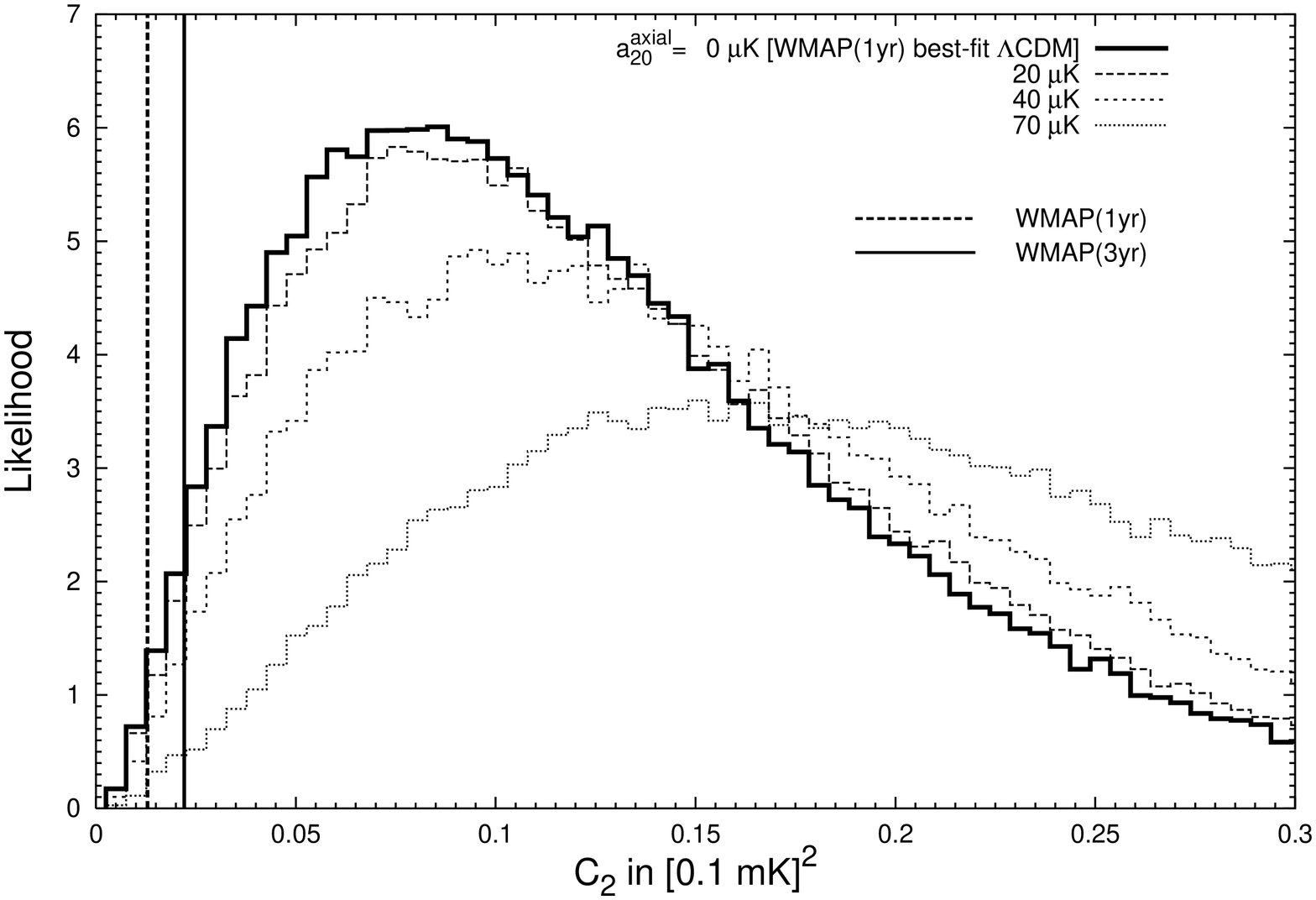,width=1.7in,angle=270}
\hspace*{4pt}
\epsfig{figure=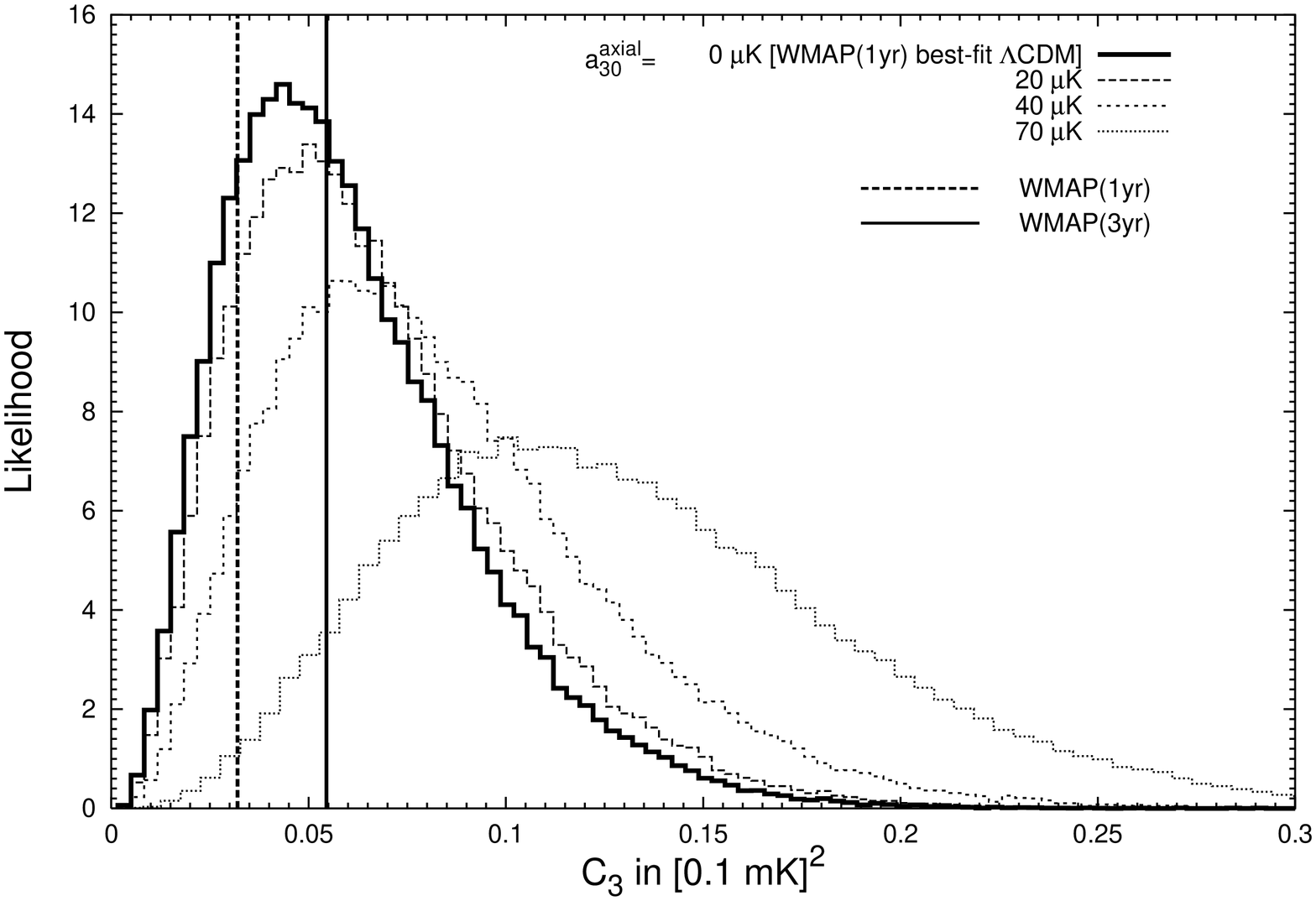,width=1.7in,angle=270}
  \caption{Likelihood of quadrupole and octopole power for increased axial
    contributions. Vertical lines denote experimental data: WMAP(1yr) cut--sky
    and WMAP(3yr) maximum likelihood estimate. Considering the quadrupole adding
    {\it any} multipole power was excluded at $>99\%$ C.L. w.r.t WMAP(1yr) but
    it is possible to add up to $60\mu$K within the same exclusion level
    w.r.t. the WMAP(3yr) value. The octopole is more resistant against axial
    contaminations as it is possible to add a whole $100\mu$K before reaching
    the same exclusion level w.r.t the updated WMAP data.}
\label{fig1}
\end{figure*}

\section{Multipole Analysis}
We study how maps of the CMB are affected by the anisotropy induced by additional 
axisymmetric contributions $a_{\ell 0}^{\rm{axial}}$ added to the quadrupole and 
octopole by using Monte Carlo (MC) methods.  
As predicted by the simplest inflationary models, we assume that the 
$a_{\ell m}$ are fully 
characterised only by angular power, for which we use the values from the best fit 
$\Lambda$CDM temperature spectrum to the WMAP data\cite{lambda}. We produced $10^5$ 
MC realisations of $\ell=2$ and $\ell=3$ for the statistical analysis.
 
The angular power spectrum is estimated
by $C_\ell = 1/(2\ell+1) \sum_m |a_{\ell m}|^2$. In fig. \ref{fig1} we 
show how the histograms for the quadrupole and octopole power compare with the 
measured values from WMAP(1yr,3yr). Considering the WMAP(1yr) cut--sky, adding 
{\it any} power to the quadrupole was already excluded at $>99\%$ C.L. whereas 
the WMAP(3yr) data allows for adding up to $a_{2 0}^{\rm{axial}}=60\mu$K in 
order to reach the same exclusion level. The octopole is quite robust against 
axial contaminations as it lies better on the fit: in order to reach the same 
exclusion level of $>99\%$ C.L. it is necessary to add $a_{3 0}^{\rm{axial}}=80\mu$K 
w.r.t. the WMAP(1yr) cut--sky and a whole $a_{3 0}^{\rm{axial}}=100\mu$K w.r.t. the 
WMAP(3yr) value. 
Considering only the WMAP(3yr) maximum likelihood estimate and increasing the
effect of local structures up to $a_{\ell
  0}^{\rm{axial}}=70\mu$K leads to an exclusion of $99.5\%$ C.L. for $C_2$ and
$92.9\%$ C.L. for $C_3$.   

The next question is what kind of phase pattern the contribution $a_{\ell
 0}^{\rm{axial}}$ will induce on the CMB sky. Using the multipole vector
formalism\cite{mvec_form} a (temperature) multipole on a sphere can be
alternatively decomposed as: 
\begin{equation}
 T_\ell = \sum_{m=-\ell}^\ell a_{\ell m} Y_{\ell m}(\theta, \varphi) = A^{(\ell)}
 \left[ \, \prod_{i=1}^\ell \left( \hat{\boldsymbol v}^{(\ell,i)} \cdot
 \hat{\boldsymbol e}(\theta, \varphi) \right) - \mathcal{L}_{\ell}(\theta, \varphi) \,
 \right]  \; \; ,
\label{mvect}
\end{equation}
where $\hat{\boldsymbol e}(\theta, \varphi) = (\sin\theta\cos\varphi ,
\sin\theta\sin\varphi , \cos\theta)$ is a radial unit vector. 
With the decomposition (\ref{mvect})
it is possible to obtain an unique factorisation of a multipole into a scalar
part $A^{(\ell)}$ which measures its total power and
$\ell$ unit vectors $\hat{\boldsymbol v}^{(\ell,i)}$ that contain all the
directional information. The signs of the multipole vectors can be absorbed into
the scalar quantity $A^{(\ell)}$, and are thus unphysical.

\begin{figure*}[b!]
\epsfig{figure=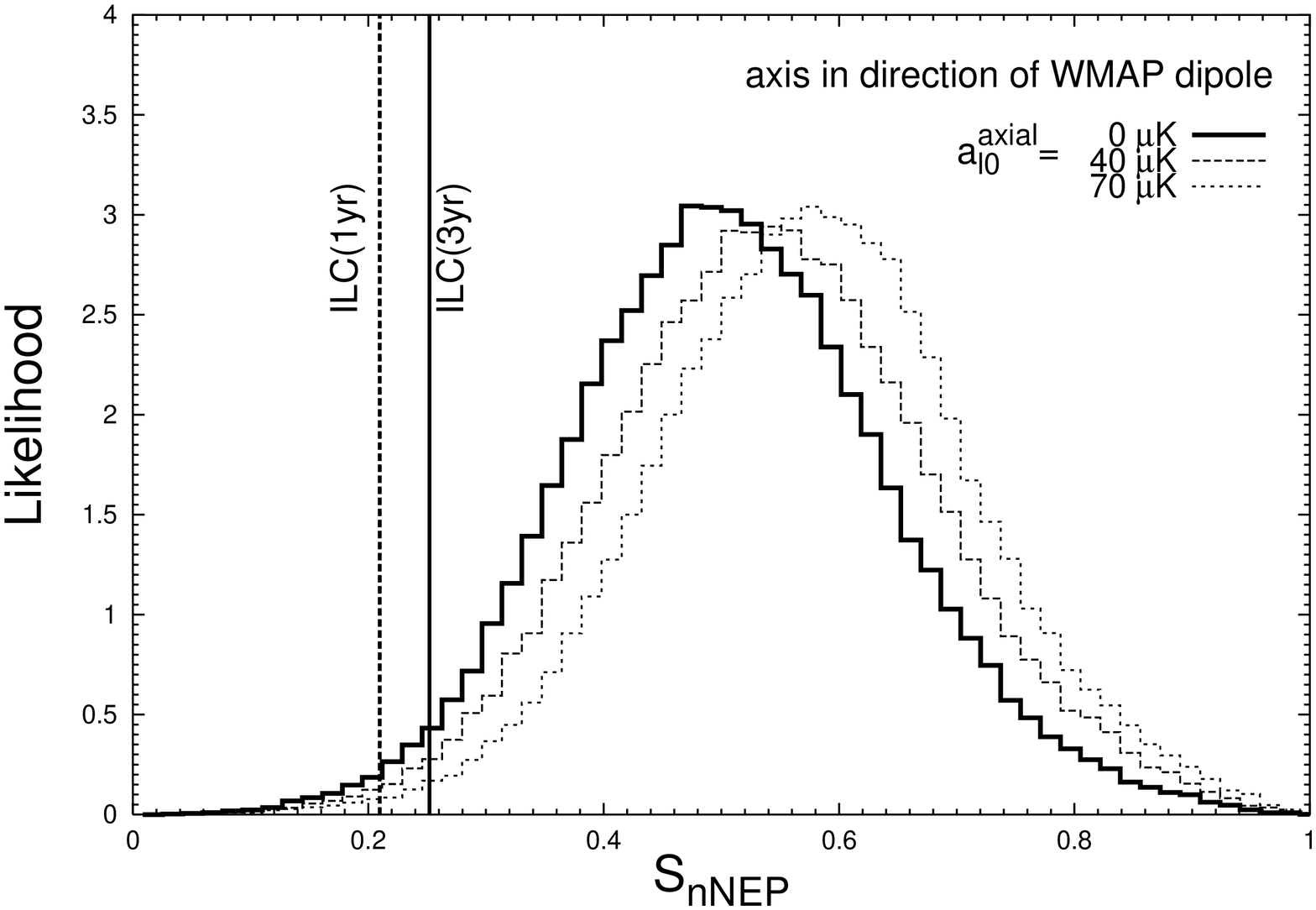,width=1.15in,angle=270}
\epsfig{figure=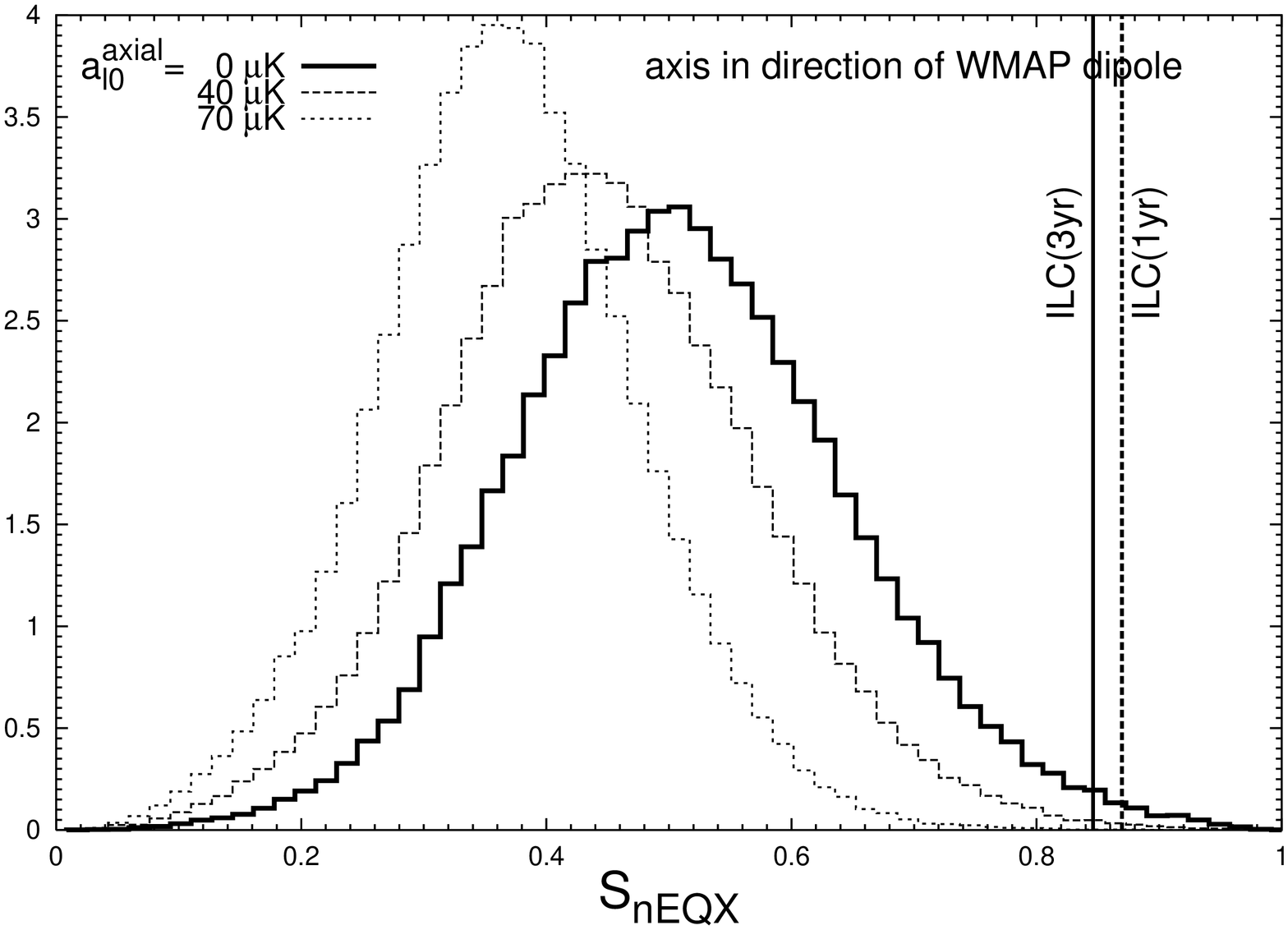,width=1.15in,angle=270}
\epsfig{figure=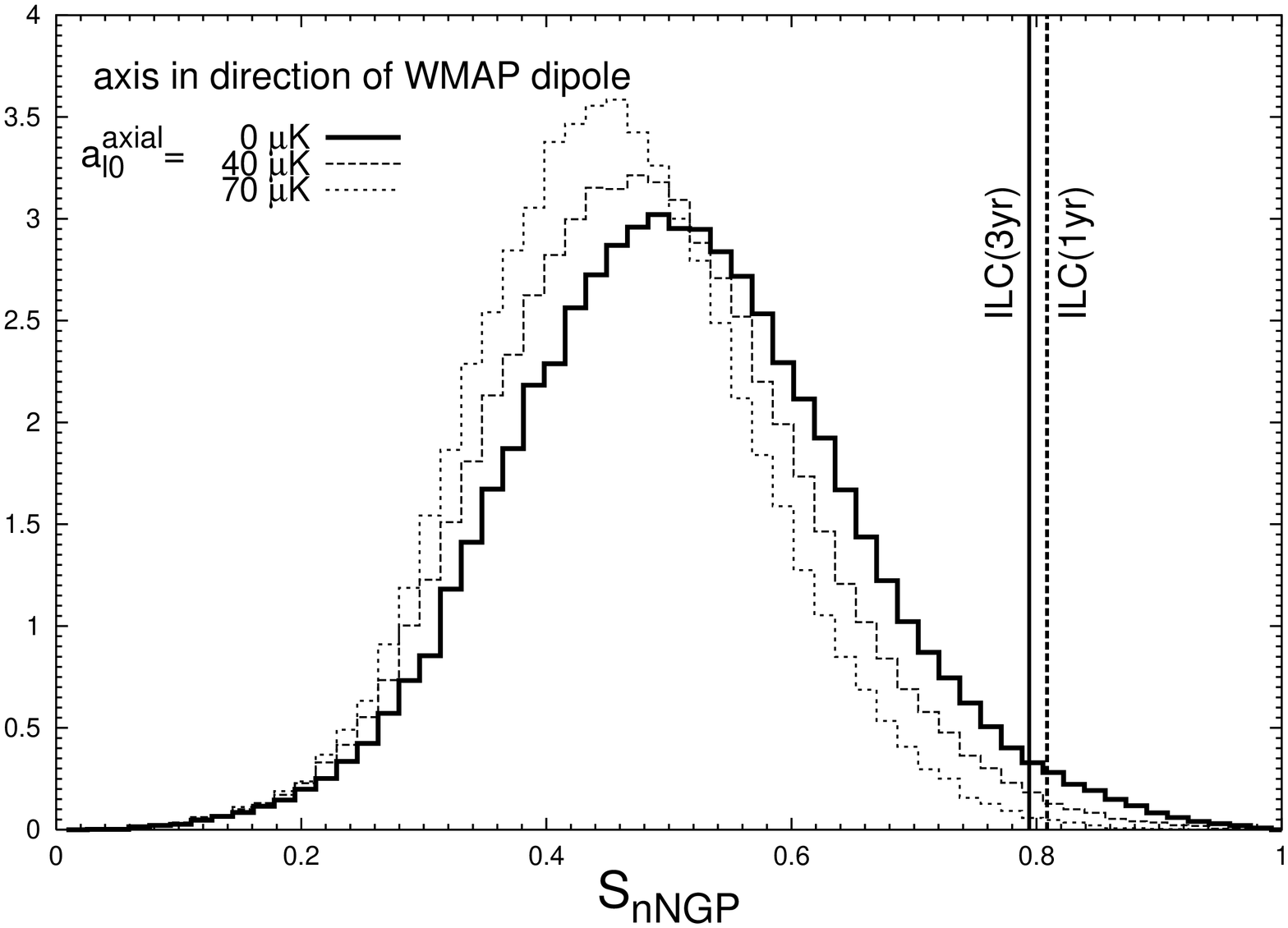,width=1.15in,angle=270}\\
\epsfig{figure=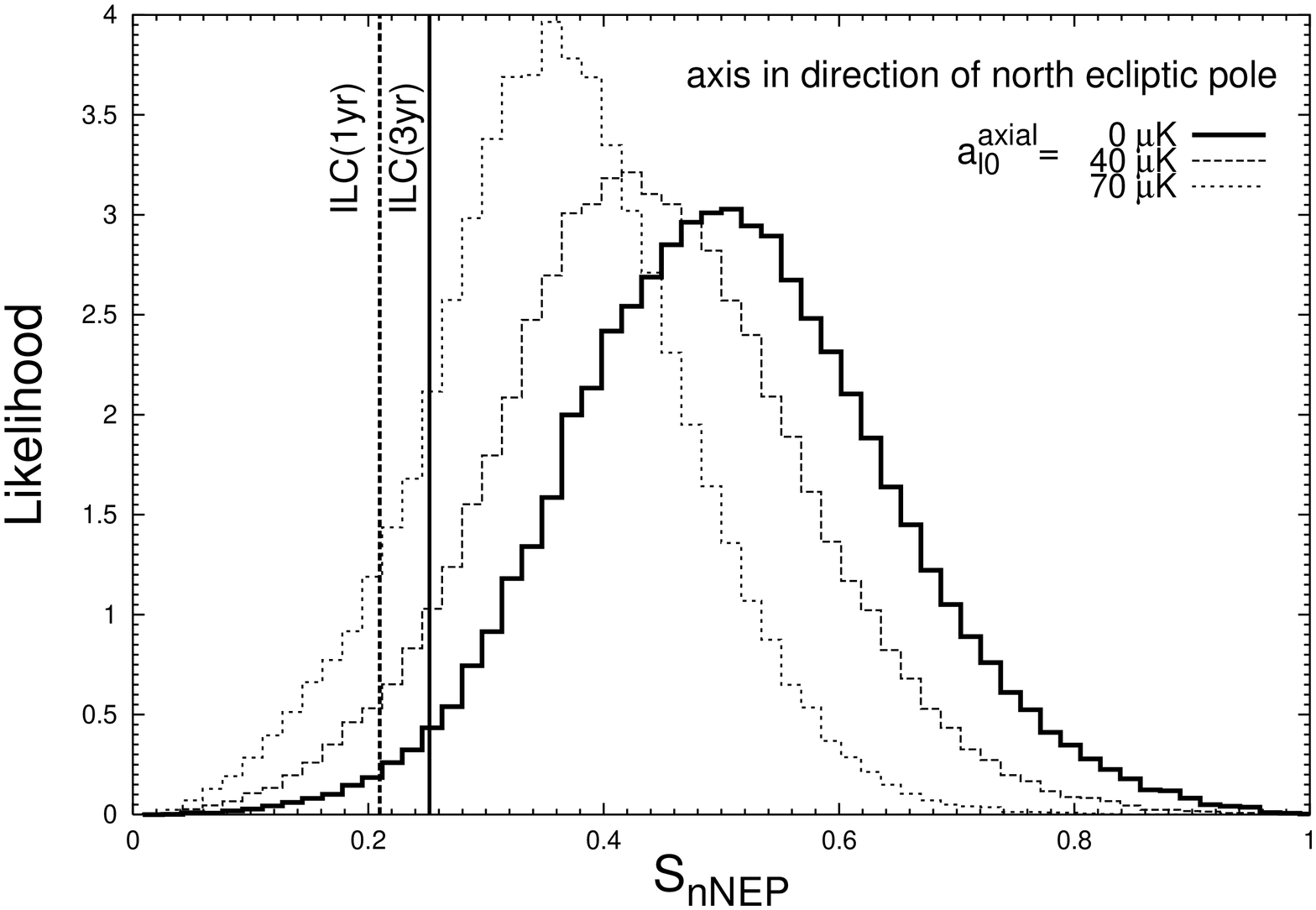,width=1.15in,angle=270}
\epsfig{figure=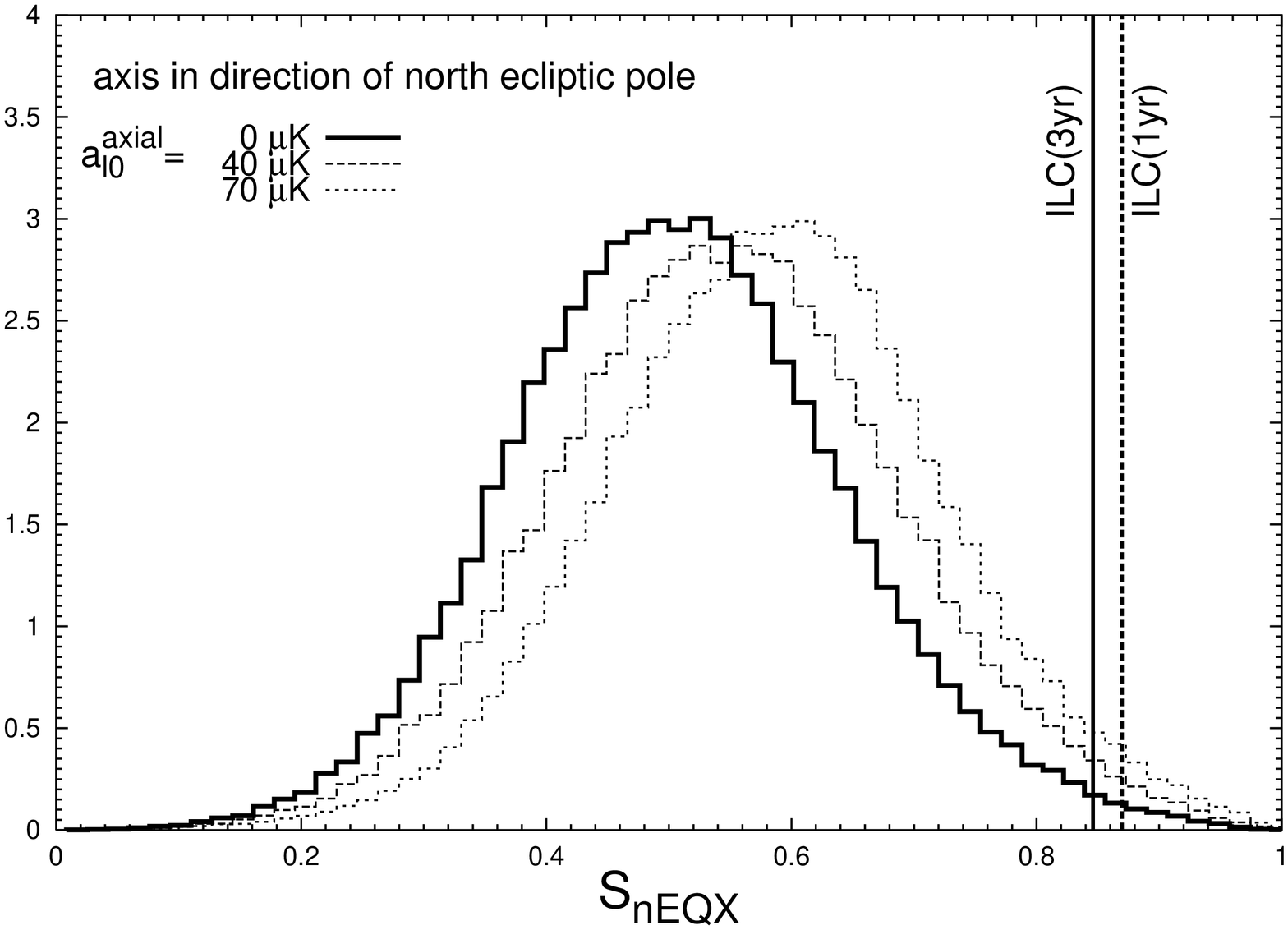,width=1.15in,angle=270}
\epsfig{figure=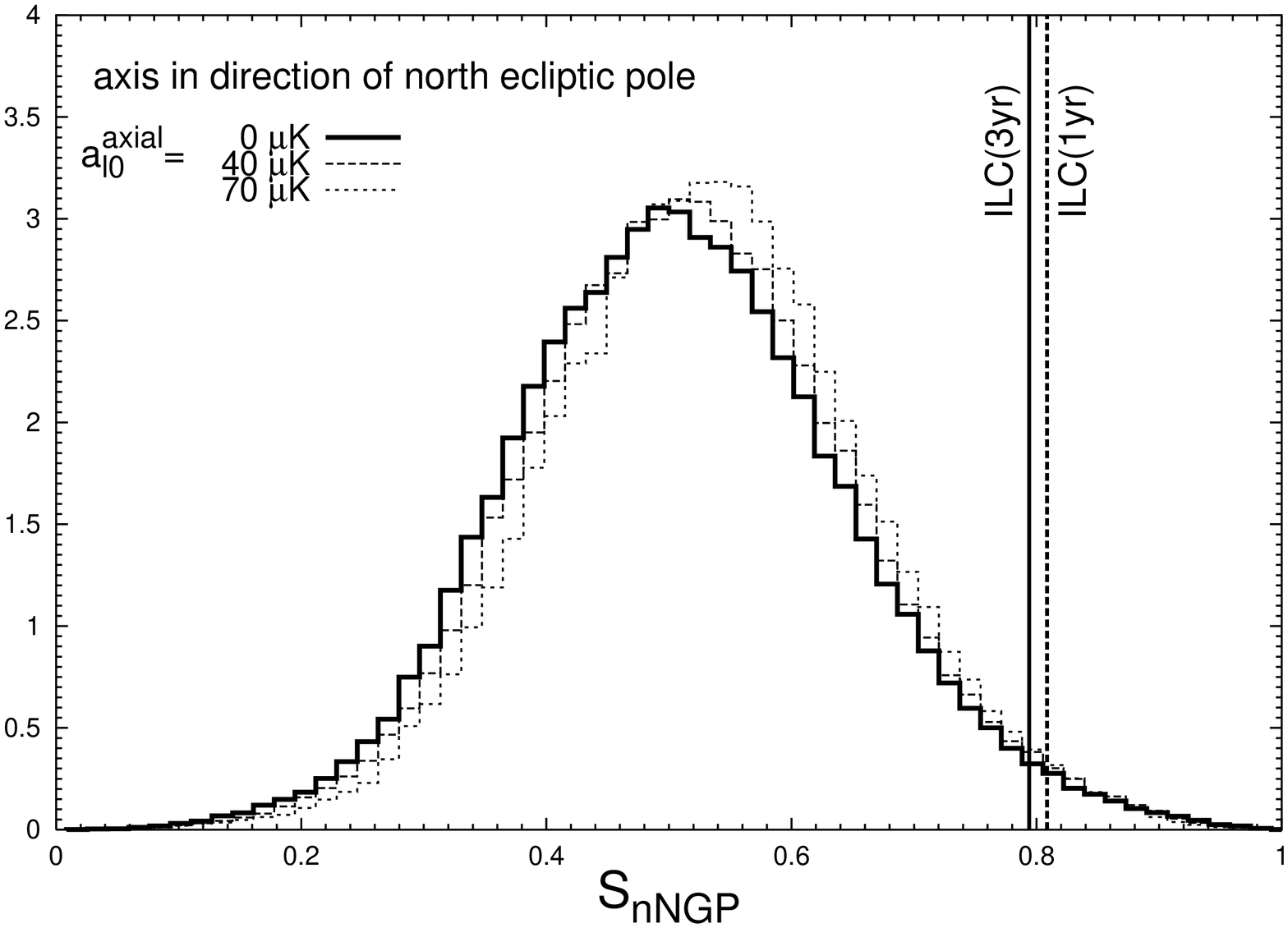,width=1.15in,angle=270}
  \caption{WMAP one- and three--year ILC maps compared to the alignment
    (\ref{eq_snx}) of quadrupole and octopole normals with physical
    directions (NEP, EQX, NGP in columns) for two orthogonal realisations of the
    preferred direction $\hat{\boldsymbol z}$ (WMAP dipole, NEP in rows). The
    bold histograms represent statistically isotropic and gaussian
    skies. Increasing the axial contribution makes the anomalies
    worse for $\hat{\boldsymbol z}=\rm{WMAP \, dipole}$, but with the exclusions being less
    significant for the ILC(3yr) than for the ILC(1yr). At the same time a Solar
    system effect is preferred by the data.} 
\label{fig2}
\end{figure*}

Introducing the $\ell (\ell -1)/2$ oriented areas $ {\boldsymbol n}^{(\ell;i,j)}
\equiv \hat{\boldsymbol v}^{(\ell,i)} \times \hat{\boldsymbol v}^{(\ell,j)}/
|\hat{\boldsymbol v}^{(\ell,i)} \times \hat{\boldsymbol v}^{(\ell,j)}|$ we are ready
to define a statistic in order to probe alignment of the normals ${\boldsymbol
  n}^{(\ell;i,j)}$ with a given physical direction $\hat{\boldsymbol x}$:
\begin{equation}
 S_{\rm{{\boldsymbol nx}}} \; \equiv \; \frac{1}{4} \, \sum_{\ell=2,3} 
 \sum_{i < j} \left| {\boldsymbol n}^{(\ell;i,j)} \cdot \hat{{\boldsymbol x}} \right| \, . 
 \label{eq_snx}
\end{equation}
\indent We test for alignment with three natural directions $\hat{{\boldsymbol x}}$: 
the north ecliptic pole (NEP), EQX and the north galactic pole 
(NGP). The results of the correlation analysis are shown in fig. \ref{fig2}:
in the first row the preferred direction $\hat{\boldsymbol z}$ coincides with
the direction of local motion, the dipole\cite{wmap1}. Here the anomaly becomes worse
when increasing the amplitude of the axial contribution. But for $\hat{{\boldsymbol x}}
=\rm{NEP}$ the exclusion becomes somewhat milder; e.g. $a_{\ell 0}^{\rm{axial}}=40\mu$K
leads to an exclusion of $99.2\%$ C.L. for ILC(1yr) but only $98.2\%$ C.L. for the
updated ILC map. Finding an alignment with the EQX though is strongly excluded
at $>99.2\%$C.L. even with a vanishing axial contribution for both one-- and
three--year data.

In the second row of fig. \ref{fig2} we let the preferred direction point to
the NEP as a complementary test. Here the probability to find
an ecliptic alignment becomes dramatically increased: with 
a $a_{\ell 0}^{\rm{axial}}=70\mu$K it is $17\%$ and $10\%$ for the ILC(3yr)
and ILC(1yr) values respectively. Regarding the three--year data the probability
for finding an EQX alignment increases from $1\%$ to $3\%$ for
$a_{\ell 0}^{\rm{axial}}=70\mu$K. The alignment with the NGP remains quite
stable for both tested directions of $\hat{\boldsymbol z}$.

\section{Conclusion}
Recent astrophysical data cataloguing our neighborhood in the
X--ray band point us to the existence of massive non--linear structures like the SSC at distances
$\sim$~100~$h^{-1}$Mpc. Besides its significant contribution to the dipole
velocity profile such a structure is able to induce anisotropies $\sim 10^{-5}$
via its RS effect. Regarding CMB modes, the spherical symmetry (LTB) which
we use to approximate the local
superstructure reduces to an axial symmetry along the line connecting our
position and the centre of the superstructure where we locate the SSC. We
produced statistically isotropic and gaussian MC maps of the CMB and computed their
$S$--statistics (\ref{eq_snx}) for alignment with generic astrophysical directions like
the NEP, EQX and NGP. The additional zonal harmonics have been added with 
increasing strength (see ref.\cite{web} for full--sky maps). 
When gauging the preferred axis to the direction of local
motion (WMAP dipole) the consistency of the data with theory becomes even worse,
albeit with less significance w.r.t. WMAP(3yr). On the
other hand an orthogonally directed (Solar system) effect would be more consistent
with the three--year data. 

Considering extended local foregrounds Abramo et al.\cite{abramo} recently
proposed that a cold spot in the direction of the local Supercluster could account
for the cross alignments of quadrupole and octopole. The cold spot would be
realised by the SZ effect of CMB photons scattering of the hot intracluster gas. 
On the other hand Inoue and Silk\cite{silk} suggest a certain geometrical
pattern of two identical voids to account for the cross alignment as well as for
the octopole planarity via the RS effect. Each of the latter approaches alone
is not fully satisfactory. Nevertheless a combined approach enfolding the RS
effect as well as the SZ effect from extended foregrounds
seems promising for the future.     
Moreover, since the local RS effect can contribute up to $10^{-5}$ to
the temperature anisotropies on large angular scales, a detailed study
is important for cross--correlating CMB data (including upcoming Planck data)
with astrophysical observations on the local large--scale structure.

\section*{Acknowledgment}
\small{It is a pleasure to thank the organisers of the 11th Marcel Grossmann
  meeting for their effort and the opportunity to speak. We acknowledge the use
  of the Legacy Archive for Microwave Background Data Analysis (LAMBDA) provided
  by the NASA Office of Space Science. The work of AR is supported by the DFG
  grant GRK 881.}


\begin{thebibliography}{99}
%
\bibitem{wmap3} G. Hinshaw et al., astro-ph/0603451.
%
\bibitem{lambda} WMAP data products at {\tt http://lambda.gsfc.nasa.gov/}
%
\bibitem{Copi2} C. Copi, D. Huterer, D. J. Schwarz and G. Starkman,
  astro-ph/0605135.   
%
%
%
\bibitem{schwarz} D. J. Schwarz, G. D. Starkman, D. Huterer and C. J. Copi, {\it
  PRL} {\bf 93}, 221301 (2004).
%
\bibitem{eriksen} H. K. Eriksen, F. K. Hansen, A. J. Banday, K. M. G\'orski and
  P. B. Lilje, {\it ApJ} {\bf 605}, 14 (2004); (Erratum) {\bf 609}, 1198 (2004).
%
\bibitem{tegmark} A. de Oliveira--Costa, M. Tegmark, M. Zaldarriaga and
  M. Hamilton, {\it Phys. Rev.} {\bf D 69}, 063516 (2004);
  A. de Oliveira--Costa and M. Tegmark, astro-ph/0603369.
%
%
\bibitem{mvec_form} C. J. Copi, D. Huterer and G. D. Starkman,
{\it Phys. Rev.} {\bf D 70}, 043515 (2004).
%
%
%
\bibitem{bianchi} T. Ghosh, A. Hajian and T. Souradeep, astro-ph/0604279. 
%
\bibitem{jaffe} T. R. Jaffe, A. J. Banday, H. K. Eriksen, K. M. Gorski and
  F. K. Hansen, astro-ph/0606046 and references therein.
%
\bibitem{alnes} H. Alnes and M. Amarzguioui, astro-ph/0607334.
%
\bibitem{moffat} J. W. Moffat JCAP {\bf 0510}, 012 (2005).
%
\bibitem{rakic} A. Raki\'c, S. R\"as\"anen and D. J. Schwarz, 
{\it MNRAS} {\bf 369}, L27 (2006).
%
\bibitem{tomita} K. Tomita {\it Phys. Rev.} {\bf D 72}, 043526 (2005),
   {\it Phys. Rev.} {\bf D 72} 103506; (Erratum) {\bf D 73} 029901.
%
\bibitem{frisch} P. C. Frisch, {\it ApJ} {\bf 632}, L143 (2005).
%
\bibitem{vale} C. Vale, astro-ph/0509039.
%
\bibitem{cooray} A. Cooray and N. Seto, JCAP {\bf 0512}, 004 (2005).
%
\bibitem{abramo} L. R. Abramo and L. Sodr\'e Jr., astro-ph/0312124;\\
  L. R. Abramo, L. Sodr\'e Jr. and C. A. Wuensche, astro-ph/0605269.
%
\bibitem{hansen} F. K. Hansen, E. Branchini, P. Mazzotta, P. Cabella and
  K. Dolag, {\it MNRAS} {\bf 361}, 753 (2005).
%
\bibitem{silk} K. T. Inoue and J. Silk, {\it ApJ} {\bf 648}, 23 (2006);
  K. T. Inoue and J. Silk, astro-ph/0612347.
%
%
%
%
\bibitem{luminet} A. Riazuelo, J. Weeks, J. P. Uzan, R. Lehoucq and
  J. P. Luminet, {\it Phys. Rev.} {\bf D 69}, 103518 (2004);
J. P. Luminet, J. Weeks, A. Riazuelo, R. Lehoucq and J. P. Uzan, {\it Nature}
{\bf 425}, 593 (2003).
%
\bibitem{cornish} J. Shapiro Key, N. J. Cornish, D. N. Spergel and
  G. D. Starkman, astro-ph/0604616; 
N. J. Cornish, D. N. Spergel, G. D. Starkman and E. Komatsu, {\it
  Phys. Rev. Lett.} {\bf 92}, 201302 (2004).
%
%
%
\bibitem{boyanovsky} D. Boyanovsky, H. J. de Vega and N. G. Sanchez,
  astro-ph/0607508; astro-ph/0607487.
%
\bibitem{campanelli} L. Campanelli, P. Cea and L. Tedesco, astro-ph/0606266.
%
\bibitem{contaldi1} C. R. Contaldi, M. Peloso, L. Kofman and A. Linde, {\it
    JCAP} {\bf 0307}, 002 (2003).
%
\bibitem{ferrer} F. Ferrer, S. R\"as\"anen and J. V\"aliviita, {\it JCAP} {\bf
    0410}, 010 (2004).
%
\bibitem{gordon} C. Gordon and W. Hu, {\it Phys. Rev.} {\bf D 70}, 083003
  (2004).
%
\bibitem{contaldi2} A. E. G\"umr\"uk\c{c}\"uo\u{g}lu, C. R. Contaldi and
  M. Peloso, astro-ph/0608405.
%
\bibitem{wolunglee} C.--H. Wu, K.--W. Ng, W. Lee, D.--S. Lee and Y.--Y. Charng,
  astro-ph/0604292.  
%
%
%
\bibitem{hofmann} S. Hofmann and O. Winkler, gr-qc/0411124.
%
\bibitem{tsujikawa} S. Tsujikawa, P. Singh and R. Maartens, {\it
    CQG} {\bf 21}, 5767 (2004).
%
%
%
\bibitem{kocevski04} D. D. Kocevski, C. R. Mullis, H. Ebeling, {\it ApJ} {\bf
    608}, 721 (2004).
%
\bibitem{kocevski05} D. D. Kocevski, H. Ebeling, {\it ApJ} {\bf 645}, 1043
  (2006).
%
\bibitem{hudson}
  M. J. Hudson, R. J. Smith, J. R. Lucey, E. Branchini, {\it MNRAS} {\bf 352},
  61 (2004).
%
\bibitem{lucey}
  J. Lucey, D. Radburn--Smith, M. Hudson, astro-ph/0412329.
%
\bibitem{proust} D. Proust et al., {\it A\&A} {\bf 447}, 133 (2006).
%
\bibitem{reessciama}
  M. J. Rees, D. W. Sciama, {\it Nature} {\bf 217}, 511 (1968). 
%
\bibitem{panek} M. Panek, {\it ApJ} {\bf 388}, 225 (1992).  
%
\bibitem{wmap1} C. L. Bennett et al., {\it ApJS} {\bf 148}, 1 (2003).
%
\bibitem{web} Full--sky maps at {\tt http://www.physik.uni-bielefeld.de/cosmology/rs.html}
%
\end{thebibliography}
\end{document}